\documentstyle[preprint,prb,aps]{revtex}
\begin{document}
\title{Conformal Anomaly and Critical Exponents  of the $XY$-Ising model}

\author{ M.P. Nightingale }
\address{Department of Physics, University of Rhode Island, \\
Kingston, Rhode Island, 02881. }
\author{E. Granato }
\address{ Laborat\'orio Associado de Sensores e Materiais, \\
Instituto Nacional de Pesquisas Espaciais, \\
12 225 S\~ao Jos\'e dos Campos, S\~ao Paulo, Brazil.}
\author{J.M. Kosterlitz }
\address{ Department of Physics, Brown University, \\
Providence, Rhode Island, 02902.}
\maketitle
\begin{abstract}

We use extensive Monte Carlo transfer matrix calculations on infinite
strips of widths $L$ up to 30 lattice spacing and a finite-size scaling
analysis to obtain critical exponents and  conformal anomaly number
$c$  for the two-dimensional $XY$-Ising model. This model is expected
to describe the critical behavior of a class of systems with
simultaneous $U(1)$ and $Z_2$ symmetries of which the fully frustrated
$XY$ model is a special case.  The effective values obtained for $c$
show a  significant decrease with $L$ at different points along the
line where the transition to the ordered phase takes place in a single
transition.  Extrapolations based on power-law corrections give values
consistent with $c=3/2$ although  larger values can not be ruled out.
Critical exponents are obtained more accurately and are consistent with
previous Monte Carlo simulations suggesting new critical behavior and
with recent calculations for the frustrated $XY$ model.

\end{abstract}
\newpage
\section{Introduction}

Recently, the critical behavior of the two-dimensional $XY$-Ising
model, consisting of $XY$ and Ising models coupled through their energy
densities, has been studied in some detail \cite{gkln91,lgk91}.  The
model is expected to describe the critical behavior of a class of
systems with $U(1)$ and $Z_2$ symmetries which includes, for example,
two-dimensional fully frustrated $XY$ (FF$XY$) models
\cite{jv77,tj83a,ms84,cd85,yd85,gk86,eg87,tk88,tk90,lgk91,jose92,%
gn93,knops94}, or alternatively, two-dimensional arrays of Josephson
junctions\cite{tj83b},  one-dimensional ladders of Josephson junctions
with charging effects \cite{eg92}, helical $XY$ models \cite{yd85} and
some surface solid-on-solid models \cite{mdn85,mdn92}. The $XY$-Ising
model is also of great theoretical interest in its own right as the
phase diagram obtained by Monte Carlo simulations gave rise to
interesting and unusual critical behavior \cite{gkln91}. Recent work by
Knops {\it et al.} \cite{knops94} has further justified the relation
between $XY$-Ising and FF$XY$ models by showing that the phase coupling
across chiral domains in the FF$XY$ model is irrelevant at
criticality.  In the subspace of parameters of the model where the $XY$
and Ising coupling constants have the same magnitude, separate $XY$ and
Ising transitions, first-order transitions and a critical line with
simultaneous $XY$ and Ising ordering were found.

The numerical study revealed that starting at the branch point where
separate $XY$ and Ising transitions merge, the line of single
transitions has a segment of continuous transitions which eventually
become first order as one moves away from the branch point. Along the
segment of continuous transitions, the critical exponents associated
with the Ising-like order parameter were found to be significantly
different from the pure Ising values and, in fact, appeared to be
non-universal, varying along the line.  Besides by critical exponents,
this critical line was also characterized by its central charge, or
conformal anomaly number $c$.  The central charge was estimated from
the finite-size scaling of the free energy of infinite strips at
criticality, obtained from Monte Carlo Transfer Matrix calculations.
The results obtained from strips of width up to $L =12$ lattice spacing
were rather surprising: the central charge appears to increase
continuously along this line, from $c\approx 1.5$ close to the branch
point to $c\approx 2$ near the tricritical point.

 Similar calculations \cite{tk90,lkg91,gn93} of the critical exponents
and central charge for the FF$XY$ model were consistent with these
results.  Although models with varying $c$ are well known, as for
example the $q$-state Potts and $O(n)$ models with a continuously
varying number of states $q$ and $n$, the behavior for $XY$-Ising model
is rather unexpected since, contrary to the previous models, the
transfer matrix can be chosen symmetric and along the critical line a
parameter is changing that does not affect the symmetry. The question
then arises if this behavior is a real effect or an artifact due to
limited strip widths.  In view of the relation between the $XY$-Ising
and  FF$XY$ models, the answer to this question may also give some
insight into the behavior of the central charge for FF$XY$ models
\cite{foda,tk90,gn93,knops94}.  Also, it is important to have improved
estimates for the critical exponents in order to be more certain about
the non-Ising nature of the critical behavior along the line of single
transitions.

In this work we report the results of extensive Monte Carlo transfer
matrix calculations, using infinite strips of widths $L$ up to 30
lattice spacing, aimed to resolve some of the issues raised by previous
calculations on the $XY$-Ising model.  Rather than attempting to
evaluate critical exponents and central charge  at several different
points along  the line of single transitions to check if these
quantities do change or remain constant along the critical line, we
have concentrated on a couple of points but performed extensive
calculations for large $L$ and used variance reduction techniques to
decrease the statistical errors.  The results for the effective value
of $c$ show a significant decrease with increasing $L$, indicating that
they even the extrapolated estimates have not yet reached their
asymptotic values for $L=30$, the largest strip width considered.
Extrapolation suggests values not inconsistent with $c=3/2$. However,
on purely numerical grounds, we can not rule out the possibility of a
larger value or even a varying $c$ along the line.  Our results for the
central charge suggest  that the recent estimates of this quantity, $c
\approx 1.6$ for the related FF$XY$ models
\cite{tk90,eg92,gn93,knops94} are likely to be subject to similar
systematic errors due to slowly decaying corrections to scaling and the
asymptotic value is in fact consistent with $c=3/2$.  The critical
exponents associated with Ising-like order parameter are obtained more
accurately, although there are some puzling inconsistencies.  The
exponents are found to be significantly different from the pure Ising
values but consistent with the previous Monte Carlo simulations which
suggested new critical behavior \cite{gkln91} and recent estimates for
the FF$XY$ model using Monte Carlo \cite{gn93} and exact numerical
transfer matrix calculations \cite{knops94}.

The paper is organized as follows. In Sec. II, we define the model and
briefly review the main features of its phase diagram, indicating the
locations near the phase boundary where the Monte Carlo transfer matrix
calculations were performed. In Sec. III we provide details on the
Monte Carlo transfer matrix method and the implementation of the
variance reduction techniques.  In Sec. IV, a finite-size scaling
analysis of the interfacial free energy is used to extract  critical
quantities. In Sec. V, we  present the numerical results for critical
couplings, exponents and central charge and in Sec. VI we discuss and
compare these results with previous calculations. Finally, Sec. VI is
devoted to the conclusions and final remarks.

\section{Model and Phase Diagram}

The $XY$-Ising model is defined by the following Hamiltonian
\cite{gkln91,lgk91}
\begin{equation}
{H\over {kT}} = - \sum_{<ij>} [(A+B s_i s_j) {\rm\bf n}_i \cdot
{\rm\bf n}_j  + C s_i s_j ],
\label{eqn.xyi}
\end{equation}
where $s = \pm 1$ is an Ising spin and ${\rm\bf n} =
(\cos\theta,\sin\theta) $ is a two-component unit vector, is an $XY$
spin. The model can be regarded as the infinite coupling limit, $h
\rightarrow \infty$, of two  $XY$ models \cite{cd85,yd85,gk86}  coupled
by a term of the form $ h\cos 2 (\theta_1 - \theta_2)$ and has a rich
phase diagram in the $A$, $B$ plane that depends strongly on the value
of $C$. The model with $A\ne B$ is relevant for the anisotropic
frustrated $XY$ model \cite{gk86} and anti-ferromagnetic restricted
solid-on-solid model \cite{mdn85}.

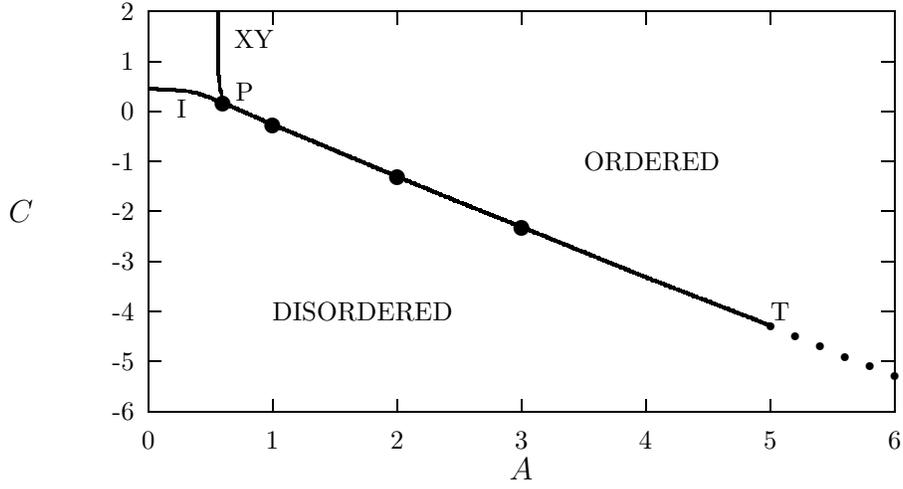
\begin{figure}
\begin{center}

%# 1 "phase_diagram.tex"
% phase diagram

% GNUPLOT: LaTeX picture
\setlength{\unitlength}{0.240900pt}
\ifx\plotpoint\undefined\newsavebox{\plotpoint}\fi
\sbox{\plotpoint}{\rule[-0.175pt]{0.350pt}{0.350pt}}%
\begin{picture}(1500,900)(0,0)
\tenrm
\sbox{\plotpoint}{\rule[-0.175pt]{0.350pt}{0.350pt}}%
\put(264,158){\rule[-0.175pt]{0.350pt}{151.526pt}}
\put(264,158){\rule[-0.175pt]{4.818pt}{0.350pt}}
\put(242,158){\makebox(0,0)[r]{-6}}
\put(1416,158){\rule[-0.175pt]{4.818pt}{0.350pt}}
\put(264,237){\rule[-0.175pt]{4.818pt}{0.350pt}}
\put(242,237){\makebox(0,0)[r]{-5}}
\put(1416,237){\rule[-0.175pt]{4.818pt}{0.350pt}}
\put(264,315){\rule[-0.175pt]{4.818pt}{0.350pt}}
\put(242,315){\makebox(0,0)[r]{-4}}
\put(1416,315){\rule[-0.175pt]{4.818pt}{0.350pt}}
\put(264,394){\rule[-0.175pt]{4.818pt}{0.350pt}}
\put(242,394){\makebox(0,0)[r]{-3}}
\put(1416,394){\rule[-0.175pt]{4.818pt}{0.350pt}}
\put(264,473){\rule[-0.175pt]{4.818pt}{0.350pt}}
\put(242,473){\makebox(0,0)[r]{-2}}
\put(1416,473){\rule[-0.175pt]{4.818pt}{0.350pt}}
\put(264,551){\rule[-0.175pt]{4.818pt}{0.350pt}}
\put(242,551){\makebox(0,0)[r]{-1}}
\put(1416,551){\rule[-0.175pt]{4.818pt}{0.350pt}}
\put(264,630){\rule[-0.175pt]{4.818pt}{0.350pt}}
\put(242,630){\makebox(0,0)[r]{0}}
\put(1416,630){\rule[-0.175pt]{4.818pt}{0.350pt}}
\put(264,708){\rule[-0.175pt]{4.818pt}{0.350pt}}
\put(242,708){\makebox(0,0)[r]{1}}
\put(1416,708){\rule[-0.175pt]{4.818pt}{0.350pt}}
\put(264,787){\rule[-0.175pt]{4.818pt}{0.350pt}}
\put(242,787){\makebox(0,0)[r]{2}}
\put(1416,787){\rule[-0.175pt]{4.818pt}{0.350pt}}
\put(264,158){\rule[-0.175pt]{0.350pt}{4.818pt}}
\put(264,113){\makebox(0,0){0}}
\put(264,767){\rule[-0.175pt]{0.350pt}{4.818pt}}
\put(459,158){\rule[-0.175pt]{0.350pt}{4.818pt}}
\put(459,113){\makebox(0,0){1}}
\put(459,767){\rule[-0.175pt]{0.350pt}{4.818pt}}
\put(655,158){\rule[-0.175pt]{0.350pt}{4.818pt}}
\put(655,113){\makebox(0,0){2}}
\put(655,767){\rule[-0.175pt]{0.350pt}{4.818pt}}
\put(850,158){\rule[-0.175pt]{0.350pt}{4.818pt}}
\put(850,113){\makebox(0,0){3}}
\put(850,767){\rule[-0.175pt]{0.350pt}{4.818pt}}
\put(1045,158){\rule[-0.175pt]{0.350pt}{4.818pt}}
\put(1045,113){\makebox(0,0){4}}
\put(1045,767){\rule[-0.175pt]{0.350pt}{4.818pt}}
\put(1241,158){\rule[-0.175pt]{0.350pt}{4.818pt}}
\put(1241,113){\makebox(0,0){5}}
\put(1241,767){\rule[-0.175pt]{0.350pt}{4.818pt}}
\put(1436,158){\rule[-0.175pt]{0.350pt}{4.818pt}}
\put(1436,113){\makebox(0,0){6}}
\put(1436,767){\rule[-0.175pt]{0.350pt}{4.818pt}}
\put(264,158){\rule[-0.175pt]{282.335pt}{0.350pt}}
\put(1436,158){\rule[-0.175pt]{0.350pt}{151.526pt}}
\put(264,787){\rule[-0.175pt]{282.335pt}{0.350pt}}
\put(45,472){\makebox(0,0)[l]{\shortstack{ $C$ }}}
\put(850,68){\makebox(0,0){ $A$ }}
\put(401,661){\makebox(0,0)[l]{P}}
\put(1241,315){\makebox(0,0)[l]{T}}
\put(309,634){\makebox(0,0)[l]{I}}
\put(399,744){\makebox(0,0)[l]{XY}}
\put(948,551){\makebox(0,0)[l]{ORDERED }}
\put(459,315){\makebox(0,0)[l]{DISORDERED}}
\put(264,158){\rule[-0.175pt]{0.350pt}{151.526pt}}
\sbox{\plotpoint}{\rule[-0.500pt]{1.000pt}{1.000pt}}%
\put(371,698){\rule[-0.500pt]{1.000pt}{21.380pt}}
\put(372,688){\rule[-0.500pt]{1.000pt}{2.349pt}}
\put(373,678){\rule[-0.500pt]{1.000pt}{2.349pt}}
\put(374,669){\rule[-0.500pt]{1.000pt}{2.349pt}}
\put(375,664){\rule[-0.500pt]{1.000pt}{1.084pt}}
\put(376,660){\rule[-0.500pt]{1.000pt}{1.084pt}}
\put(377,655){\rule[-0.500pt]{1.000pt}{1.084pt}}
\put(378,651){\rule[-0.500pt]{1.000pt}{1.084pt}}
\put(379,646){\rule[-0.500pt]{1.000pt}{1.084pt}}
\put(380,642){\rule[-0.500pt]{1.000pt}{1.084pt}}
\put(264,665){\usebox{\plotpoint}}
\put(264,665){\rule[-0.500pt]{3.553pt}{1.000pt}}
\put(278,664){\rule[-0.500pt]{3.553pt}{1.000pt}}
\put(293,663){\rule[-0.500pt]{3.553pt}{1.000pt}}
\put(308,662){\rule[-0.500pt]{3.553pt}{1.000pt}}
\put(323,661){\rule[-0.500pt]{1.144pt}{1.000pt}}
\put(327,660){\rule[-0.500pt]{1.144pt}{1.000pt}}
\put(332,659){\rule[-0.500pt]{1.144pt}{1.000pt}}
\put(337,658){\rule[-0.500pt]{1.144pt}{1.000pt}}
\put(342,657){\usebox{\plotpoint}}
\put(344,656){\usebox{\plotpoint}}
\put(347,655){\usebox{\plotpoint}}
\put(350,654){\usebox{\plotpoint}}
\put(353,653){\usebox{\plotpoint}}
\put(356,652){\usebox{\plotpoint}}
\put(359,651){\usebox{\plotpoint}}
\put(362,650){\usebox{\plotpoint}}
\put(363,649){\usebox{\plotpoint}}
\put(365,648){\usebox{\plotpoint}}
\put(367,647){\usebox{\plotpoint}}
\put(369,646){\usebox{\plotpoint}}
\put(370,645){\usebox{\plotpoint}}
\put(374,644){\usebox{\plotpoint}}
\put(377,643){\usebox{\plotpoint}}
\put(381,642){\usebox{\plotpoint}}
\put(383,641){\usebox{\plotpoint}}
\put(385,640){\usebox{\plotpoint}}
\put(387,639){\usebox{\plotpoint}}
\put(390,638){\usebox{\plotpoint}}
\put(392,637){\usebox{\plotpoint}}
\put(394,636){\usebox{\plotpoint}}
\put(397,635){\usebox{\plotpoint}}
\put(399,634){\usebox{\plotpoint}}
\put(401,633){\usebox{\plotpoint}}
\put(403,632){\usebox{\plotpoint}}
\put(406,631){\usebox{\plotpoint}}
\put(408,630){\usebox{\plotpoint}}
\put(410,629){\usebox{\plotpoint}}
\put(413,628){\usebox{\plotpoint}}
\put(415,627){\usebox{\plotpoint}}
\put(417,626){\usebox{\plotpoint}}
\put(420,625){\usebox{\plotpoint}}
\put(422,624){\usebox{\plotpoint}}
\put(424,623){\usebox{\plotpoint}}
\put(426,622){\usebox{\plotpoint}}
\put(429,621){\usebox{\plotpoint}}
\put(431,620){\usebox{\plotpoint}}
\put(433,619){\usebox{\plotpoint}}
\put(436,618){\usebox{\plotpoint}}
\put(438,617){\usebox{\plotpoint}}
\put(440,616){\usebox{\plotpoint}}
\put(442,615){\usebox{\plotpoint}}
\put(445,614){\usebox{\plotpoint}}
\put(447,613){\usebox{\plotpoint}}
\put(449,612){\usebox{\plotpoint}}
\put(452,611){\usebox{\plotpoint}}
\put(454,610){\usebox{\plotpoint}}
\put(456,609){\usebox{\plotpoint}}
\put(459,608){\usebox{\plotpoint}}
\put(461,607){\usebox{\plotpoint}}
\put(463,606){\usebox{\plotpoint}}
\put(466,605){\usebox{\plotpoint}}
\put(468,604){\usebox{\plotpoint}}
\put(470,603){\usebox{\plotpoint}}
\put(473,602){\usebox{\plotpoint}}
\put(475,601){\usebox{\plotpoint}}
\put(478,600){\usebox{\plotpoint}}
\put(480,599){\usebox{\plotpoint}}
\put(482,598){\usebox{\plotpoint}}
\put(485,597){\usebox{\plotpoint}}
\put(487,596){\usebox{\plotpoint}}
\put(490,595){\usebox{\plotpoint}}
\put(492,594){\usebox{\plotpoint}}
\put(494,593){\usebox{\plotpoint}}
\put(497,592){\usebox{\plotpoint}}
\put(499,591){\usebox{\plotpoint}}
\put(502,590){\usebox{\plotpoint}}
\put(504,589){\usebox{\plotpoint}}
\put(506,588){\usebox{\plotpoint}}
\put(509,587){\usebox{\plotpoint}}
\put(511,586){\usebox{\plotpoint}}
\put(513,585){\usebox{\plotpoint}}
\put(516,584){\usebox{\plotpoint}}
\put(518,583){\usebox{\plotpoint}}
\put(521,582){\usebox{\plotpoint}}
\put(523,581){\usebox{\plotpoint}}
\put(525,580){\usebox{\plotpoint}}
\put(528,579){\usebox{\plotpoint}}
\put(530,578){\usebox{\plotpoint}}
\put(533,577){\usebox{\plotpoint}}
\put(535,576){\usebox{\plotpoint}}
\put(537,575){\usebox{\plotpoint}}
\put(540,574){\usebox{\plotpoint}}
\put(542,573){\usebox{\plotpoint}}
\put(545,572){\usebox{\plotpoint}}
\put(547,571){\usebox{\plotpoint}}
\put(549,570){\usebox{\plotpoint}}
\put(552,569){\usebox{\plotpoint}}
\put(554,568){\usebox{\plotpoint}}
\put(557,567){\usebox{\plotpoint}}
\put(559,566){\usebox{\plotpoint}}
\put(561,565){\usebox{\plotpoint}}
\put(564,564){\usebox{\plotpoint}}
\put(566,563){\usebox{\plotpoint}}
\put(568,562){\usebox{\plotpoint}}
\put(571,561){\usebox{\plotpoint}}
\put(573,560){\usebox{\plotpoint}}
\put(576,559){\usebox{\plotpoint}}
\put(578,558){\usebox{\plotpoint}}
\put(580,557){\usebox{\plotpoint}}
\put(583,556){\usebox{\plotpoint}}
\put(585,555){\usebox{\plotpoint}}
\put(588,554){\usebox{\plotpoint}}
\put(590,553){\usebox{\plotpoint}}
\put(592,552){\usebox{\plotpoint}}
\put(595,551){\usebox{\plotpoint}}
\put(597,550){\usebox{\plotpoint}}
\put(600,549){\usebox{\plotpoint}}
\put(602,548){\usebox{\plotpoint}}
\put(604,547){\usebox{\plotpoint}}
\put(607,546){\usebox{\plotpoint}}
\put(609,545){\usebox{\plotpoint}}
\put(611,544){\usebox{\plotpoint}}
\put(614,543){\usebox{\plotpoint}}
\put(616,542){\usebox{\plotpoint}}
\put(619,541){\usebox{\plotpoint}}
\put(621,540){\usebox{\plotpoint}}
\put(623,539){\usebox{\plotpoint}}
\put(626,538){\usebox{\plotpoint}}
\put(628,537){\usebox{\plotpoint}}
\put(631,536){\usebox{\plotpoint}}
\put(633,535){\usebox{\plotpoint}}
\put(635,534){\usebox{\plotpoint}}
\put(638,533){\usebox{\plotpoint}}
\put(640,532){\usebox{\plotpoint}}
\put(643,531){\usebox{\plotpoint}}
\put(645,530){\usebox{\plotpoint}}
\put(647,529){\usebox{\plotpoint}}
\put(650,528){\usebox{\plotpoint}}
\put(652,527){\usebox{\plotpoint}}
\put(655,526){\usebox{\plotpoint}}
\put(657,525){\usebox{\plotpoint}}
\put(659,524){\usebox{\plotpoint}}
\put(662,523){\usebox{\plotpoint}}
\put(664,522){\usebox{\plotpoint}}
\put(667,521){\usebox{\plotpoint}}
\put(669,520){\usebox{\plotpoint}}
\put(671,519){\usebox{\plotpoint}}
\put(674,518){\usebox{\plotpoint}}
\put(676,517){\usebox{\plotpoint}}
\put(679,516){\usebox{\plotpoint}}
\put(681,515){\usebox{\plotpoint}}
\put(684,514){\usebox{\plotpoint}}
\put(686,513){\usebox{\plotpoint}}
\put(688,512){\usebox{\plotpoint}}
\put(691,511){\usebox{\plotpoint}}
\put(693,510){\usebox{\plotpoint}}
\put(696,509){\usebox{\plotpoint}}
\put(698,508){\usebox{\plotpoint}}
\put(701,507){\usebox{\plotpoint}}
\put(703,506){\usebox{\plotpoint}}
\put(705,505){\usebox{\plotpoint}}
\put(708,504){\usebox{\plotpoint}}
\put(710,503){\usebox{\plotpoint}}
\put(713,502){\usebox{\plotpoint}}
\put(715,501){\usebox{\plotpoint}}
\put(718,500){\usebox{\plotpoint}}
\put(720,499){\usebox{\plotpoint}}
\put(722,498){\usebox{\plotpoint}}
\put(725,497){\usebox{\plotpoint}}
\put(727,496){\usebox{\plotpoint}}
\put(730,495){\usebox{\plotpoint}}
\put(732,494){\usebox{\plotpoint}}
\put(735,493){\usebox{\plotpoint}}
\put(737,492){\usebox{\plotpoint}}
\put(739,491){\usebox{\plotpoint}}
\put(742,490){\usebox{\plotpoint}}
\put(744,489){\usebox{\plotpoint}}
\put(747,488){\usebox{\plotpoint}}
\put(749,487){\usebox{\plotpoint}}
\put(751,486){\usebox{\plotpoint}}
\put(754,485){\usebox{\plotpoint}}
\put(757,484){\usebox{\plotpoint}}
\put(759,483){\usebox{\plotpoint}}
\put(762,482){\usebox{\plotpoint}}
\put(764,481){\usebox{\plotpoint}}
\put(767,480){\usebox{\plotpoint}}
\put(769,479){\usebox{\plotpoint}}
\put(772,478){\usebox{\plotpoint}}
\put(774,477){\usebox{\plotpoint}}
\put(777,476){\usebox{\plotpoint}}
\put(779,475){\usebox{\plotpoint}}
\put(782,474){\usebox{\plotpoint}}
\put(784,473){\usebox{\plotpoint}}
\put(787,472){\usebox{\plotpoint}}
\put(789,471){\usebox{\plotpoint}}
\put(792,470){\usebox{\plotpoint}}
\put(794,469){\usebox{\plotpoint}}
\put(797,468){\usebox{\plotpoint}}
\put(799,467){\usebox{\plotpoint}}
\put(802,466){\usebox{\plotpoint}}
\put(804,465){\usebox{\plotpoint}}
\put(807,464){\usebox{\plotpoint}}
\put(809,463){\usebox{\plotpoint}}
\put(812,462){\usebox{\plotpoint}}
\put(814,461){\usebox{\plotpoint}}
\put(817,460){\usebox{\plotpoint}}
\put(819,459){\usebox{\plotpoint}}
\put(822,458){\usebox{\plotpoint}}
\put(824,457){\usebox{\plotpoint}}
\put(827,456){\usebox{\plotpoint}}
\put(829,455){\usebox{\plotpoint}}
\put(832,454){\usebox{\plotpoint}}
\put(834,453){\usebox{\plotpoint}}
\put(837,452){\usebox{\plotpoint}}
\put(839,451){\usebox{\plotpoint}}
\put(842,450){\usebox{\plotpoint}}
\put(844,449){\usebox{\plotpoint}}
\put(847,448){\usebox{\plotpoint}}
\put(849,447){\usebox{\plotpoint}}
\put(852,446){\usebox{\plotpoint}}
\put(854,445){\usebox{\plotpoint}}
\put(857,444){\usebox{\plotpoint}}
\put(859,443){\usebox{\plotpoint}}
\put(862,442){\usebox{\plotpoint}}
\put(864,441){\usebox{\plotpoint}}
\put(867,440){\usebox{\plotpoint}}
\put(869,439){\usebox{\plotpoint}}
\put(872,438){\usebox{\plotpoint}}
\put(874,437){\usebox{\plotpoint}}
\put(877,436){\usebox{\plotpoint}}
\put(879,435){\usebox{\plotpoint}}
\put(882,434){\usebox{\plotpoint}}
\put(884,433){\usebox{\plotpoint}}
\put(887,432){\usebox{\plotpoint}}
\put(889,431){\usebox{\plotpoint}}
\put(891,430){\usebox{\plotpoint}}
\put(894,429){\usebox{\plotpoint}}
\put(896,428){\usebox{\plotpoint}}
\put(899,427){\usebox{\plotpoint}}
\put(901,426){\usebox{\plotpoint}}
\put(904,425){\usebox{\plotpoint}}
\put(906,424){\usebox{\plotpoint}}
\put(909,423){\usebox{\plotpoint}}
\put(911,422){\usebox{\plotpoint}}
\put(914,421){\usebox{\plotpoint}}
\put(916,420){\usebox{\plotpoint}}
\put(919,419){\usebox{\plotpoint}}
\put(921,418){\usebox{\plotpoint}}
\put(924,417){\usebox{\plotpoint}}
\put(926,416){\usebox{\plotpoint}}
\put(928,415){\usebox{\plotpoint}}
\put(931,414){\usebox{\plotpoint}}
\put(933,413){\usebox{\plotpoint}}
\put(936,412){\usebox{\plotpoint}}
\put(938,411){\usebox{\plotpoint}}
\put(941,410){\usebox{\plotpoint}}
\put(943,409){\usebox{\plotpoint}}
\put(946,408){\usebox{\plotpoint}}
\put(948,407){\usebox{\plotpoint}}
\put(951,406){\usebox{\plotpoint}}
\put(953,405){\usebox{\plotpoint}}
\put(956,404){\usebox{\plotpoint}}
\put(958,403){\usebox{\plotpoint}}
\put(961,402){\usebox{\plotpoint}}
\put(963,401){\usebox{\plotpoint}}
\put(966,400){\usebox{\plotpoint}}
\put(968,399){\usebox{\plotpoint}}
\put(970,398){\usebox{\plotpoint}}
\put(973,397){\usebox{\plotpoint}}
\put(975,396){\usebox{\plotpoint}}
\put(978,395){\usebox{\plotpoint}}
\put(980,394){\usebox{\plotpoint}}
\put(983,393){\usebox{\plotpoint}}
\put(985,392){\usebox{\plotpoint}}
\put(988,391){\usebox{\plotpoint}}
\put(990,390){\usebox{\plotpoint}}
\put(993,389){\usebox{\plotpoint}}
\put(995,388){\usebox{\plotpoint}}
\put(998,387){\usebox{\plotpoint}}
\put(1000,386){\usebox{\plotpoint}}
\put(1003,385){\usebox{\plotpoint}}
\put(1005,384){\usebox{\plotpoint}}
\put(1007,383){\usebox{\plotpoint}}
\put(1010,382){\usebox{\plotpoint}}
\put(1012,381){\usebox{\plotpoint}}
\put(1015,380){\usebox{\plotpoint}}
\put(1017,379){\usebox{\plotpoint}}
\put(1020,378){\usebox{\plotpoint}}
\put(1022,377){\usebox{\plotpoint}}
\put(1025,376){\usebox{\plotpoint}}
\put(1027,375){\usebox{\plotpoint}}
\put(1030,374){\usebox{\plotpoint}}
\put(1032,373){\usebox{\plotpoint}}
\put(1035,372){\usebox{\plotpoint}}
\put(1037,371){\usebox{\plotpoint}}
\put(1040,370){\usebox{\plotpoint}}
\put(1042,369){\usebox{\plotpoint}}
\put(1045,368){\usebox{\plotpoint}}
\put(1047,367){\usebox{\plotpoint}}
\put(1050,366){\usebox{\plotpoint}}
\put(1052,365){\usebox{\plotpoint}}
\put(1055,364){\usebox{\plotpoint}}
\put(1057,363){\usebox{\plotpoint}}
\put(1060,362){\usebox{\plotpoint}}
\put(1063,361){\usebox{\plotpoint}}
\put(1065,360){\usebox{\plotpoint}}
\put(1068,359){\usebox{\plotpoint}}
\put(1070,358){\usebox{\plotpoint}}
\put(1073,357){\usebox{\plotpoint}}
\put(1075,356){\usebox{\plotpoint}}
\put(1078,355){\usebox{\plotpoint}}
\put(1081,354){\usebox{\plotpoint}}
\put(1083,353){\usebox{\plotpoint}}
\put(1086,352){\usebox{\plotpoint}}
\put(1088,351){\usebox{\plotpoint}}
\put(1091,350){\usebox{\plotpoint}}
\put(1094,349){\usebox{\plotpoint}}
\put(1096,348){\usebox{\plotpoint}}
\put(1099,347){\usebox{\plotpoint}}
\put(1101,346){\usebox{\plotpoint}}
\put(1104,345){\usebox{\plotpoint}}
\put(1106,344){\usebox{\plotpoint}}
\put(1109,343){\usebox{\plotpoint}}
\put(1112,342){\usebox{\plotpoint}}
\put(1114,341){\usebox{\plotpoint}}
\put(1117,340){\usebox{\plotpoint}}
\put(1119,339){\usebox{\plotpoint}}
\put(1122,338){\usebox{\plotpoint}}
\put(1124,337){\usebox{\plotpoint}}
\put(1127,336){\usebox{\plotpoint}}
\put(1130,335){\usebox{\plotpoint}}
\put(1132,334){\usebox{\plotpoint}}
\put(1135,333){\usebox{\plotpoint}}
\put(1137,332){\usebox{\plotpoint}}
\put(1140,331){\usebox{\plotpoint}}
\put(1143,330){\usebox{\plotpoint}}
\put(1145,329){\usebox{\plotpoint}}
\put(1148,328){\usebox{\plotpoint}}
\put(1150,327){\usebox{\plotpoint}}
\put(1153,326){\usebox{\plotpoint}}
\put(1155,325){\usebox{\plotpoint}}
\put(1158,324){\usebox{\plotpoint}}
\put(1161,323){\usebox{\plotpoint}}
\put(1163,322){\usebox{\plotpoint}}
\put(1166,321){\usebox{\plotpoint}}
\put(1168,320){\usebox{\plotpoint}}
\put(1171,319){\usebox{\plotpoint}}
\put(1173,318){\usebox{\plotpoint}}
\put(1176,317){\usebox{\plotpoint}}
\put(1179,316){\usebox{\plotpoint}}
\put(1181,315){\usebox{\plotpoint}}
\put(1184,314){\usebox{\plotpoint}}
\put(1186,313){\usebox{\plotpoint}}
\put(1189,312){\usebox{\plotpoint}}
\put(1192,311){\usebox{\plotpoint}}
\put(1194,310){\usebox{\plotpoint}}
\put(1197,309){\usebox{\plotpoint}}
\put(1199,308){\usebox{\plotpoint}}
\put(1202,307){\usebox{\plotpoint}}
\put(1204,306){\usebox{\plotpoint}}
\put(1207,305){\usebox{\plotpoint}}
\put(1210,304){\usebox{\plotpoint}}
\put(1212,303){\usebox{\plotpoint}}
\put(1215,302){\usebox{\plotpoint}}
\put(1217,301){\usebox{\plotpoint}}
\put(1220,300){\usebox{\plotpoint}}
\put(1222,299){\usebox{\plotpoint}}
\put(1225,298){\usebox{\plotpoint}}
\put(1228,297){\usebox{\plotpoint}}
\put(1230,296){\usebox{\plotpoint}}
\put(1233,295){\usebox{\plotpoint}}
\put(1235,294){\usebox{\plotpoint}}
\put(1238,293){\usebox{\plotpoint}}
\sbox{\plotpoint}{\rule[-0.175pt]{0.350pt}{0.350pt}}%
\put(1241,292){\circle*{12}}
\put(1280,276){\circle*{12}}
\put(1319,260){\circle*{12}}
\put(1358,244){\circle*{12}}
\put(1397,229){\circle*{12}}
\put(1436,213){\circle*{12}}
\put(381,642){\circle*{24}}
\put(459,608){\circle*{24}}
\put(655,526){\circle*{24}}
\put(850,447){\circle*{24}}
\end{picture}
%# 896 "paper_050595.acpp"

\caption{ Phase diagram of the $XY$-Ising model
\protect{\cite{gkln91,lgk91}}.  Solid and dotted lines indicate
continuous and first-order transitions, respectively.  Filled circles
indicate the locations where the present calculations were performed.}

\label{fig.phdiag}
\end{center}
\end{figure}

In this work we will be concerned with the critical behavior of the
$XY$-Ising model of Eq.~(\ref{eqn.xyi}), defined on a square lattice,
in the subspace $A=B$,
\begin{equation}
{H\over {kT}} = - \sum_{<ij>} [A ( 1+ s_i s_j) {\rm\bf n}_i \cdot
{\rm\bf n}_j  + C s_i s_j ],
\end{equation}
which is relevant for the isotropic frustrated $XY$ model or its
one-dimensional quantum version \cite{eg92}.  The phase diagram
obtained by Monte Carlo simulations is shown in Fig.~\ref{fig.phdiag}
and consists of three branches joining at $P$, in the ferromagnetic
region $A>0$, $ A+C > 0$. One of the branches corresponds to a single
transition with simultaneous loss of $XY$ and Ising order and the other
two to separate Kosterlitz-Thouless (KT) and Ising transitions.  An
important feature of the phase diagram is that there is no phase with
Ising disorder and $XY$ order thus indicating that Ising disorder
induces also $XY$ disorder in this model. This is related to the
special symmetry under the transformation
\begin{equation}
{\rm\bf n}_j\rightarrow s_j{\rm\bf n}_j
\label{eq.n_to_sn}
\end{equation}

which holds if $A=B$, since $XY$ spins are not coupled across an Ising
domain wall where $s_{i}s_{j} +1 =0$.  The behavior of FF$XY$ model
coresponds to the behavior of this model along a particular path
through this phase diagram. The available numerical results for the
standard FF$XY$ model\cite{gkln91,lkg91,gn93,knops94} are consistent
with a single transition but generalized versions could  correspond to
a path through the double transition region. In fact, a Coulomb-gas
representation of the FF$XY$ model with an additional coupling between
nearest neighbor vortices has a phase diagram with identical structure
\cite{tk88}.  In the one-dimensional quantum version of the frustrated
$XY$ model \cite{eg92}, related to the zero-temperature transition of
Josephson-junction ladders, double or single transitions will result
depending on the ratio between inter-chain and intra-chain couplings.
In the Monte Carlo simulations \cite{lgk91}, the critical line $PT$ in
Fig.~\ref{fig.phdiag} appears to be non-universal as the critical
exponents associated with the Ising order parameter were found to vary
systematically along this line. In addition, a preliminary evaluation
of the central charge \cite{gkln91} $c$ using data for the free energy
of infinite strips obtained from Monte Carlo transfer matrix  appeared
to indicate that $c$ varies from $c \approx 1.5$ near $P$  to $c=2$
near $T$. These results for the central charge were based on strips of
width $L$ up to 12 lattice spacings. However, this range of $L$ and the
numerical noise in the data does not allow one to extrapolate to the
large $L$ limit and these estimates are thus subject to systematic
errors. To obtain  accurate estimates it is necessary to perform
calculations for larger $L$ and also to reduce the errors.  Rather than
attempting  to evaluate $c$ at several different points along the line
$PT$ in order to check if this quantity changes or remains constant
along the line, we have concentrated our attention at a few points but
performed extensive calculations for large $L$ and used variance
reduction techniques to decrease the errors.  The calculations
discussed in the following sections were performed primarily along the
cuts through the critical line as indicated in Fig.~\ref{fig.phdiag}.

\section{Monte Carlo Transfer Matrix}

Estimates of the free energy density per lattice site was computed
using the Monte Carlo transfer matrix method.  We give a brief summary
of the essentials of this method and refer the reader to
Ref.~\onlinecite{Nightingale.Privman.Review} for more details.

Helical boundary conditions are convenient for these computations.  In
this case the transfer matrix can be chosen to be a sparse matrix for
the case where one matrix multiplication corresponds to the addition of
one surface site to the lattice, as illustrated in
Fig.~\ref{fig.transfer_matrix}.  The sparseness follows from the fact
that from any given configuration of surface sites only those new
configurations can be reached that differ at that newly added site
only.  We used a transfer matrix defined by
\begin{eqnarray}
& T(s_1,\dots,s_L;{\rm\bf m}_1 ,\dots,{\rm\bf m}_L\,|\,
t_1,\dots,t_L;{\rm\bf n}_1 ,\dots,{\rm\bf n}_L)=\\ \nonumber
& e ^ { (A\,{\rm\bf n}_{L-1} \cdot {\rm\bf n}_L +
B\,{\rm\bf n}_{L-1} \cdot {\rm\bf n}_L t_{L-1} t_L +
C t_{L-1} t_L
+A\,{\rm\bf n}_L \cdot {\rm\bf m}_1 +
B\,{\rm\bf n}_L \cdot {\rm\bf m}_L s_L s_1 +
C s_{L-1} s_L) }
\prod_{k=2}^L \delta({\rm\bf m}_k-{\rm\bf n}_{k-1}) \delta_{s_k,t_{k-1}}.
\label{eqn.transfermatrix}
\end{eqnarray}

The statistical variance of transfer matrix Monte Carlo computations is
proportional to the variance of the quantity
\begin{equation}
\mu=\sum_{s_1} \int d{\rm/bf
m}_1\,T(s_1,\dots,s_L;{\rm/bf m}_1,\dots,{\rm\bf m}_L\,|\,
  t_1,\dots,t_L;{\rm\bf n}_1,\dots,{\rm\bf n}_L).
\label{eq.mu}
\end{equation}

The variance can be reduced by applying the transfer matrix algorithm
to a similarity transform $\hat T$ of the transfer matrix $T$ defined
in Eq.~(\ref{eqn.transfermatrix}).  The transformation requires an
optimized trial eigenvector $\psi_{\rm T}$ and is defined as follows:
\begin{equation}
\hat T(\{s\};\{{\rm\bf m}\}\,|\,\{t\};\{{\rm\bf n}\})=\psi_{\rm
T}(\{s\};\{{\rm\bf m}\})T(\{s\};\{{\rm\bf m}\}\,|\,\{t\};\{{\rm\bf n}\})
/\psi_{\rm T}(\{t\};\{{\rm\bf n}\})
\end{equation}
where $\{x\} \equiv x_{1},\dots,x_{L}$.

In the ideal case, when $\psi_{\rm T}$ is an exact left eigenvector,
the local eigenvalue $\hat\mu$, defined by Eq.~(\ref{eq.mu}) with $T$
replaced by $\hat T$, is a constant - an eigenvalue of the transfer
matrix.  In practice, the better the quality of the trial function, the
smaller the statistical noise in the Monte Carlo estimates of the
transfer matrix eigenvalues.

For the design of good trial states it is helpful to realize that the
dominant left eigenvector is proportional to the conditional partition
function of a semi-infinite lattice, extending to infinity towards the
left, as indicated in Fig.~\ref{fig.transfer_matrix}, conditional on
the microscopic state of the surface.

Our computations used the following form for the trial vectors:
\begin{equation}
\psi_{\rm T} (s_1,\dots,s_L;{\rm\bf n}_1,\dots,{\rm\bf n}_L)=
\exp\left(\sum_{i,j}^\ast
\left( A_{i,j}\,{\rm\bf n}_i \cdot {\rm\bf n}_j +
B_{i,j}\, {\rm\bf n}_i \cdot {\rm\bf n}_j s_i s_j+ C_{i,j}
s_i s_j\right)\right) .
\label{eqn.trial.xy.ising}
\end{equation}
Here the parameters $A_{ij}$, $B_{ij}$, and $C_{ij}$ are variational
parameters, which are chosen so as to minimize the variance of $\hat
\mu$, as described in detail in
Ref.~\onlinecite{Nightingale.Privman.Review}.  In the expression
(\ref{eqn.trial.xy.ising}) the asterisk indicates that the sum over the
pairs of surface sites labeled $i$ and $j$ is truncated, as is required
for for computational efficiency.  To truncate in a way that allows
systematic improvement of the quality of the trial function, it is
necessary to guess for which pairs of sites $i$ and $j$ the interaction
parameters $A_{i,j}$ in Eq.~(\ref{eqn.trial.xy.ising}) have the largest
magnitudes, and similarly for $B_{i,j}$, and $C_{i,j}$ .  Obviously,
interaction strengths will decay with distance, but owing to the
helical boundary conditions and the surface defect, the geometrical
distance is not quite correct.  Instead, a distance can be defined
between surface sites $i$ and $j$ of the semi-infinite strip
(illustrated in Fig.~\ref{fig.transfer_matrix}) as the length of the
shortest path that:  {\it (a)} connects sites $i$ and $j$; {\it (b)}
passes only through bulk sites (indicated by full circles in
Fig.~\ref{fig.transfer_matrix}); and {\it (c)} travels along the edges
of the square lattice.

The reason for excluding surface sites from the path is that the
correlations described by the interaction parameters $A_{ij}$, {\it
etc.} are mediated only via bulk sites, since those are the only ones
that contain variables that are not frozen in the conditional partition
function.  Fig.~\ref{fig.transfer_matrix} shows a path of length
three.  Owing to the presence of the surface defect no two paths are
strictly equivalent and, since the surface interactions can be regarded
as functions of the minimal path defined above, all parameters have to
be assumed different.  However, the transformed transfer matrix $\hat
T$ depends only on the ratio of the values of two trial states shifted
by a single lattice unit along the surface. By artificially imposing
translational invariance on the interaction parameters, one can produce
cancellations in the computation. This reduces the number of arithmetic
operations from order $L$ (in the absence of translation symmetry) to a
number of the order of the maximum path length at which the interaction
are truncated.

Suppose that interactions in the trial function are truncated at path
length $l$, measured in units of the lattice spacing, then the
following compromise seems to work satisfactorily: give those
interaction parameters the same values that are {\it(a)\,} farther away
from the defect than $l$, and {\it(b)\,} would be equivalent by
translation symmetry in the case of simple periodic boundary
conditions.  In particular, this means that all interactions associated
with paths that cross the defect are allowed to be different in the
computation.  It should be noted that this approximation can only be
improved to a point: as soon as many-body interactions appear that are
of greater strength than pair interactions included in the trial
function, ignoring the many-body interactions makes it impossible and
pointless to determine the two-body interactions.

We are interested only in studying the behavior of systems with $B=C$,
but the twisted boundary conditions force us also to consider the case
where $B=-C$, which is obtained by inverting either the Ising or the
$XY$ variables on one sub-lattice.  In all of these cases we used trial
functions in which the corresponding relation was maintained between
the interaction parameters appearing in the trial vector,
Eq.~(\ref{eqn.trial.xy.ising}), i.e., $B_{ij}=C_{ij}$ if $i$ and $j$
belong to the same  sub-lattice and $B_{ij}=-C_{ij}$ otherwise or if
twisted boundary conditions are used.

As a final comment we mention that by using the variance reduction
scheme mentioned above the Monte Carlo calculation can be accelerated
roughly by a factor of two hundred
\cite{Nightingale.Bloete.unpublished}.

\section{Finite-size scaling }

To locate the critical couplings and determine the critical exponents
we will do a finite-size scaling analysis of interfacial free energies.
Since the model contains both $XY$ and Ising variables, there are in
principle two types of interfacial free energies that can be determined
by a suitable choice of the boundary conditions. If a twist in the
Ising variables is imposed by anti-periodic boundary conditions, a
domain wall is forced along the infinite strip  and the associated
interfacial free energy can be obtained from the difference per surface
unit the free energies of systems with periodic and anti-periodic
boundary conditions. On the other hand, if the same procedure is
followed for the $XY$ variables, a smooth phase twist of $\pi$ is
forced across the infinite strip. The transfer matrix calculations are
done for an $L \times \infty$ strip with $L$ even and helical boundary
conditions. With this set-up, it is simple to introduce independent
twists in the Ising and $XY$ degrees of freedom.

The interfacial free energy of an Ising domain wall of length $L$
along the strip is given by
\begin{equation}
\Delta F_{\rm I} = L^2 [f(A,A,C) - f(A,-A,-C)],
\label{eqn.dfi}
\end{equation}
where $f(A,B,C)$ is the free energy per site   of the $XY$-Ising model
of Eq.~(\ref{eqn.xyi}). The parameters $A$ and $C$ are chosen so that
the ground state is ferromagnetic, $A>0$ and $A+C > 0$, so that taking
$B \rightarrow -B = -A$ induces a domain wall between the two
anti-ferromagnetic Ising ground states. Similarly, a twist of $\pi$ in
the $XY$ degrees of freedom is induced by $A \rightarrow -A$ and $B
\rightarrow -B$ so that
\begin{equation}
\Delta F_{XY} = L^2[f(A,A,C)-f(-A,-A,C)],
\label{eqn.dfxy}
\end{equation}
and the helicity modulus $\gamma $ is given by
\begin{equation}
\gamma = 2 \Delta F_{XY} / \pi^2 .
\label{eqn.helicity}
\end{equation}

At a conventional second-order transition, the interracial
free energy has the simple scaling form
\begin{equation}
\Delta F(A,C;L) = {\cal A}(L^{y_T} t),
\label{eqn.scaling}
\end{equation}
where ${\cal A}(u)$ is a scaling function and $t(A,C)$ is the
non-linear scaling field measuring the distance from the critical
point; the thermal scaling exponent $y_T$ is related by $\nu=1/y_T$ to
the exponent $\nu$, which describes the divergence of the correlation
length at criticality. In our analysis, we fix one of of the
parameters, $A$ or $C$, and expand $t$ to quadratic order, i.e., for
fixed $C$ we have $t = A-A_c(C) + k [A-A_c(C)]^2$ and similarly when
$A$ is kept fixed.  In the vicinity of the critical coupling $t=0$, a
standard finite size scaling expansion in $u=t L^{y_T}$ for $u\approx
0$ yields
\begin{equation}
\Delta F(A,C;L) = a_o + a_1 u + a_2 u^2 + \cdots.
\label{eqn.expansion}
\end{equation}
With our convention, $u$ is positive in the ordered phase,  $\Delta F$
will increase with $L$ for $u >0$, decrease for $u <0$ and be a
constant at $u=0$ for sufficiently large $L$ so that corrections to
scaling have become negligible.

Sufficiently close to $u=0$,  Eq.~(\ref{eqn.expansion}) can be used to
obtain accurate estimates of the critical exponent $y_T$ (or
equivalently $\nu$) and the critical values $A$ and $C$. The expansion
is truncated at some high order ($u^5$ in some cases).  A critical
dimension $x^{\rm (d)}$ of a disorder operator can be obtained from the
constant $a_0$ via
\begin{equation}
x^{\rm (d)}={a_0\over 2 \pi}.
\end{equation}
The critical dimension $x$ describes the decay with distance $r$ at
criticality of the two-point correlation function $g(r)$ of an operator
determined by the choice of boundary conditions:  $g(r) \propto
r^{-2x^{\rm (d)}}$. The scaling exponent $y^{\rm (d)}=2-x^{\rm (d)}$
describes the behavior of this operator under scaling.

As mentioned above, we consider two kinds of anti-periodic boundary
conditions.  Subscripts will be used to distinguish the exponents
$x^{\rm (d)}$ and $y^{\rm (d)}$ of the associated operators.  In the
case of anti-periodic boundary conditions in the $XY$ variables, the
conjugate operator is a vortex of strength ${1 \over 2}$, measured in
units $2\pi$. Such an operator is represented as the end point of a
path on the dual lattice: $XY$ bonds crossing this path have their
interactions changed from $A$ to $-A$.  Because of the symmetry of the
model under the transformation given in Eq.~(\ref{eq.n_to_sn}) this
operator is equivalent to one where the $B$ is changed to $-B$.  The
exponents of the $1 \over 2$-vortex will be denoted by $x_{XY,{1 \over
2}}^{(d)}$ and $y_{XY,{1 \over 2}}^{(d)}$.  The operator corresponding
to the case of antisymmetric boundary conditions for the Ising
variables is the standard Ising disorder operator. The exponents for
this case with be denoted $x^{\rm (d)}_{\rm I}$ and $y^{\rm (d)}_{\rm
I}$.  For self-dual models or models for which a renormalization
mapping to the Gaussian model exists, the disorder operators can be
related to order operators, but we cannot derive either of those
properties for the $XY$-Ising model.

The critical exponents were estimated by making a scaling plot of
$\Delta F$  in which the parameters were estimated by a constrained
least-square fits with the critical couplings fixed at their most
reliable estimates, i.e., those obtained by extrapolation from the
Ising domain wall data. Unfortunately, the discontinuity in the
helicity modulus $\gamma$ is not accessible by similar finite size
scaling considerations since the discontinuity in $\gamma$ is defined
in the thermodynamic limit $L \rightarrow \infty$ and
\begin{equation}
\Delta\gamma = 2[{\cal A}(\infty)-{\cal A}(-\infty)]/\pi^{2}.
\label{eqn.dischelicity}
\end{equation}
A rough estimate from Fig.~\ref{ScalingPlotxia1.xytwist} for ${\cal
A}(\pm \infty)$ gives $\Delta \gamma \approx 1.3$ which is about double
the value $2/\pi$ of the two-dimensional $XY$ model. This estimate is
not very reliable but we can say with a considerable degree of
confidence that $\Delta \gamma$ is much larger than $2/\pi$ in this
coupled $XY$-Ising model and in the FF$XY$ model.

In addition to critical exponents, another important quantity which
provides information on the nature of the critical behavior is the
central charge $c$ of the conformal invariance.  This quantity can be
obtained from the amplitude of the singular part of the free energy per
site \cite{bcn86}, at criticality, in the infinite strip by
\begin{equation}
f(A_c,C_c,L) \approx f_\infty + {{\pi c}\over {6 L^2}}
\label{eqn.c}
\end{equation}
for sufficiently large $L$, where $f_\infty$ denotes the regular
contribution to the free energy at the critical point.  The central
charge classifies the possible conformally invariant critical
behaviors. For example, for the pure Ising model, $c=1/2$, and along
the critical line of the $XY$ model $c=1$. Although $c$ is only defined
at criticality, Eq.~(\ref{eqn.c}) can be used to define a size and
coupling dependent effective central charge $c(A,C,L)$ away from the
critical point. If this quantity is now identified as the function
$c(g)$  defined in the $c$-theorem of Zamolodchikov \cite{zam86}, where
$g$ stands for a set of coupling constants, this quantity should have a
well-defined behavior near criticality since, according to the
$c$-theorem, $c(g)$ is a monotonically decreasing function under a
renormalization group transformation and reaches a constant value equal
to the central charge at the fixed point. An interesting consequence of
this identification is that $c(A,C,L)$ should have a maximum near the
critical line of single transitions in the $XY$-Ising model with a
lower bound $c(A,C,L)\ge 3/2$ and away from the critical line should
converge to either $c(A,C,L) = 1$ in the $XY$ ordered phase or to
$c(A,C,L) = 0$ in the remaining phases. Our calculations are consistent
with this behavior but we found that the maximum in $c(A,C,L)$ does not
provide an accurate procedure to locate the critical couplings since it
is rather flat within a wide range of couplings near $u=0$. To obtain
an estimate of the central charge $c$ at criticality we first
accurately locate the critical couplings using the non-linear fitting
of Eq.~(\ref{eqn.expansion}) and extract a size-dependent $c(L)$ from
the singular part of the free energy in Eq.~(\ref{eqn.c}), which is
subsequently extrapolated to $L \to \infty$.

\section{Estimates of critical points, exponents and conformal anomalies}

We computed eigenvalues of the transfer matrix for various points along
the critical curve  and used these to extract estimates for the central
charge.  In two cases we recomputed the critical points themselves from
a scaling analysis of the interface free energy and helicity modulus.
We start our discussion with the latter.
Fig.~\ref{ScalingPlotxia1.istwist} is a scaling plot of the Ising
interface free energy as a  function of $A$ at fixed $C=0.2885$.
Fig.~\ref{ScalingPlotxia1.xytwist} is the same for $XY$ interface,
obtained by choosing boundary conditions that induce a twist of $\pi$
in the $XY$ variables. Figs.~\ref{ScalingPlotxia2.istwist} and
\ref{ScalingPlotxia2.xytwist} are analogous plots for the case $A=2$
with varying $C$.  The scaling plots for the systems with anti-periodic
boundary conditions in the Ising variables do not display statistically
significant deviations from the scaling hypothesis.  However, this is
not the case for the scaling plots for systems with a twist in the $XY$
variables as shown most clearly by Fig.~\ref{ScalingPlotxia1.xytwist}.
In fact, significant changes are ,e.g. in the critical amplitude, are
observed in the ``scaling plot'' if smaller system sizes are omitted
from the fit.

In all cases, there are strong corrections to scaling for small
systems.  This is demonstrated in Figs.~\ref{xia1.Kc_L} and
\ref{xia2.Kc_L}, plots of the estimated effective critical couplings
versus inverse system size.  The effective coupling associated with
size $L$ was obtained by a least-squares fit to system sizes including
sizes $L$ and up.

By extrapolation assuming overly conservative $1/L^2$ corrections, we
obtain the following estimated critical points:  $A=1.0014$ (Ising
twist) and $A=1.0025$ ($XY$ twist) at $C=0.2885$, where the first of
these two is presumably the more reliable one.  Similarly, for $A=2$
the results are $C=1.318$ (Ising twist) and $C=1.316$ ($XY$ twist).
Our estimates for the critical exponents are summarized in Table
\ref{tab.exponents}; the plots in Figs.~\ref{xia1.Kc_L} and
\ref{xia2.Kc_L} may serve to provide rough error bars.

Finally, we estimated the conformal anomaly $c$ along the critical line
using Eq.~(\ref{eqn.c}) and taking $2, 3$ or  $4$ consecutive system
sizes. This defines an effective $c(L)$ at $L^{-1}=(L_{\rm
min}^{-1}+L_{\rm max}^{-1})/2 $, where $L_{\rm min}$ and $L_{\rm max}$
are the smallest and largest size used in the fit. Critical points not
mentioned above, {\it viz.} ($A=3,\ C=-2.3250$) and
($A=0.6,\ C=0.1520$), were taken from the estimates provided in Refs.
\onlinecite{gkln91} and \onlinecite{lgk91}.  The results are summarized
in Fig.~\ref{c.plot}.

\section{Discussion}

The results obtained from the Ising interface free energy, summarized
in Table~\ref{tab.exponents}, seem sufficiently accurate to exclude
pure Ising critical exponents ($y_T=1$ and $y^{(\rm d)}_{\rm I}=15/8$)
for the point $A \approx 1$.  Our numerical results agree with those
for the 19-vertex model obtained by Knops {\it et al.}\cite{knops94},
who find $y_T=1.23(3)$ and $y^{(\rm d)}_{\rm I}=1.80(1)$. Within the
sizable uncertainties in the estimates of the same exponents for $A =
2$, we find no evidence for variation of these exponents along the
critical line.  The results obtained for the thermal exponent $y_T$ are
consistent with those from direct Monte Carlo simulations \cite{lgk91}
of the $XY$-Ising model: 1.19(4) for $A=1$ and $y_T=1.18(4)$ for
$A=2$.  We note, however, that the latter computations indicate
variation along the critical line of the scaling exponent of the order
parameter, an exponent which was not computed in the present transfer
matrix Monte Carlo approach.  Lee {\it et al.} found:  $y^{(o)}_{\rm
I}=1.85(2)$ for $A=1$ and $y^{(o)}_{\rm I}=1.80(2)$.

There is a serious internal inconsistency in our estimates for $y_T$ as
obtained from the the Ising interface and those obtained from the $XY$
interface.  Although Figs. \ref{xia1.yt_L} and \ref{xia2.yt_L} display
strong corrections to scaling, there is no indication that the two ways
of computing this thermal exponent will become consistent in the limit
of large systems.  This calls in question the validity of the basic
assumption of scaling theory, {\it viz.,} that there is a single
divergent length scale in this system as the critical point is
approached along a temperature-like direction.

The results for the critical exponents $y_T$ and  $y_I^{(d)}$ for the
$XY$-Ising model are consistent with similar Monte Carlo transfer matrix
calculations  for the FF$XY$ model on a square lattice \cite{gn93}
($y_T=1.25(6)$, $y_I^{(d)}=1.81(2)$ ) and its one-dimensional quantum
version \cite{eg92} ($y_T=1.24(6)$, $y_I^{(d)}=1.77(2)$), although the
strip widths are much larger here and the accuracy much better.
Estimates of the critical exponent $y_T$ for the FF$XY$ model  obtained
from Monte Carlo simulations \cite{lgk91}, $1.21(3)$ (square lattice)
and $1.18(3)$ (triangular lattice) are also in good agreement with the
result for the $XY$-Ising model and seem to support an $XY$-Ising
universality class for these systems.

The exponent $y_I^{(o)}=1.85$ for the $XY$-Ising model obtained from
Monte Carlo simulations \cite{lgk91} is significantly larger than the
result for $y_I^{(d)}$ in Table I. This discrepancy is also observed in
the results for the FF$XY$ model \cite{gn93,knops94} and can   in part
be attributed to corrections due to the effective free boundary
conditions for the $XY$ degrees of freedom at criticality. As argued in
the context of the FF$XY$ model \cite{knops94}, since the $XY$ degrees of
freedom are uncoupled across an Ising domain wall, the $XY$ variables
should be regarded as having free boundary conditions instead of
periodic ones. This results in a correction to the estimate of $x^{(d)}
\rightarrow x^{(d)} - 1/16$ which seems to improve the agreement
between $y_I^{(o)}$ and $y_I^{(d)}$ although, as mentioned in Sec. IV,
the precise relation between these exponents is not known.

The results for the exponents in Table I and the scaling plots for
${\cal A}(u)$ for the interface free energies are based on the naive
assumption that the $XY$ and Ising correlation lengths behave as
$\xi_{\alpha} \sim t^{-1/y_{\alpha}}$ with $y_{\alpha}$ determined by
independent best fits for the Ising and $XY$ interfacial free
energies.  Such a procedure would certainly be incorrect if the $XY$
and Ising variables were decoupled as then $\log \xi_{XY} \sim
t^{-1/2}$. In the present case, these degrees are strongly coupled and
the appropriate scaling form is not known and the $XY$ degrees of
freedom are probably subject to large slowly decaying
corrections-to-scaling making the analysis of the data fraught with
difficulty and uncertainty. A detailed analysis of the data for the
$XY$ twist free energy shows that the estimates of the finite-size
scaling parameters $a_{i}$ of Eq.(12) are not stable and depend on the
number of data points included in the fit. For this reason, the scaling
plots of ${\cal A}(u)$ are somewhat misleading and a naive use of
Eq.(13) to estimate $x_{XY}^{(d)}$ yields the $L=4$ value for $y^{(d)}
= 2-x^{(d)}$ of Figs.(11,12). The reason behind this is that the small
$L$ data has the lowest $\chi^{2}$ and is most heavily weighted in the
scaling plots of Figs. 4 and 6. It is amusing to note that the use of
these estimates together with the with the relation, valid for a self
dual Gaussian model$^{14}$, $x_{XY,1/2}^{(d)}x_{XY,1}^{(o)}=1/16$,
gives results in agreement with those of Knops et al$^{14}$. However,
we consider this to be fortuitous and not to be taken seriously.
Another difficulty with analyzing numerical data for the $XY$ twist
free energy is that there must be a crossover to a low temperature
Gaussian fixed line when $u>>0$ as the low temperature phase must be
just a $XY$ model when there is long-range Ising order.

The well-known difficulties of analyzing numerical data for the
helicity modulus in this system are compounded by this cross-over so it
is not surprizing that we are unable to make definitive statements
about the critical exponents for the $XY$ variables. One might try
using a dual roughening representation, but there are negative
Boltzmann weights at the critical point in the dual representation
which will lead to some difficulties. Despite being able to go to
relatively large strip widths of $L=30$ we are unable to reach definite
conclusions about the critical behavior of the $XY$ degrees of freedom
except to say that our simple ansatz for the scaling of the $XY$ twist
is inadequate and corrections to scaling should be included in the
analysis, but we do not know the form these should take. Also, we are
unable to estimate the discontinuity $\pi\Delta\gamma$ in the helicity
modulus except to say that, at the critical point $\pi\gamma \approx
1.1$ and, at $T_{c}^{-}$, $\pi\Delta\gamma \approx 4$ which we believe
to be a fairly realistic estimate.

We now consider the results for the central charge in
Fig.~\ref{c.plot}.  The results for $A =0.6$ correspond to the branch
point in Fig.~\ref{fig.phdiag} as estimated from Monte Carlo
simulations \cite{lgk91}.  Convergence is very poor in this case. The
effective $c$ starts at $c = 1.5$ for small systems, decreases very
slowly for intermediate systems and them decreases rapidly for the
largest system sizes. It is not possible to estimate the asymptotic
value for this case.  In fact, this behavior suggests inaccuracy in the
estimate of the critical point.  The other curves in
Fig.~\ref{fig.phdiag} correspond to different points along the line of
single transitions. Again, corrections to scaling are decaying too
slowly as a function of system size to allow an accurate estimate of
$c$ in the large $L$ limit. However, $c=3/2$ along the line is not
inconsistent with the data. This is shown in Table \ref{tab.central}
where the central charge is estimated assuming  power-law corrections
of the form $\alpha/L^{3-s}+\beta/L^{4-s}$ the leading correction to
the free energy, $\pi c/ 6 L^2$, in Eq.~\ref{eqn.c}. We chose $s=0.2$
so that we could simultaneously fit the results for $A\approx 1$ and
$A=2$.  This value, $c=3/2$, would be the expected one if the critical
behavior along the single line could be described as a superposition of
critical Ising and $XY$ models \cite{foda}.  Our results for the
critical exponents $y_t$ and $y_h$ however are not consistent with this
hypothesis and suggest that the coupling between the Ising an $XY$
degrees of freedom is vital.  The results of the extrapolations should
also be viewed with caution since they are not completely justified.
There could be other corrections as $\exp ( - a L)$ or $\ln L /L$ but
due to the noise in the data any attempt to include such terms in the
extrapolation is pointless.

\section{Conclusions}

We have obtained critical exponents and the central charge for the
$XY$-Ising model using Monte Carlo transfer matrix calculations on
infinite strips of widths $L$ as large as $30$ lattice spacings.  The
results for $c$ show a significant decrease with increasing $L$ but
converge very slowly to an asymptotic value. An extrapolation
procedure  indicates that these values are not inconsistent with
$c=3/2$. However, on purely numerical grounds, we can not rule out the
possibility of a larger value or even $c$ varying along the line of
single transitions.  Our results for the central charge suggest  that
the recent estimates of this quantity for the related FF$XY$ models
are likely to be subject to similar systematic errors due to slowly
decaying corrections-to-scaling and the asymptotic value is consistent
with $c=3/2$.  The critical exponents associated with Ising-like order
parameter are obtained more accurately and are found to be
significantly different from the pure Ising values but are consistent
with previous Monte Carlo simulations which suggested new critical
behavior  and also with recent estimates for the FF$XY$ model using
Monte Carlo  and exact  transfer matrix calculations.

\acknowledgements

We thank H.W.J.  Bl\"ote and J. Lee for many helpful discussions. E.G.
was supported by Funda\c c\~ao de Amparo \`a Pesquisa do Estado de
S\~ao Paulo (FAPESP, Proc. No. 92/0963-5) and Conselho Nacional de
Desenvolvimento Cient\'ifico e Tecnol\'ogico (CNPq). J.M.K. was
supported by the NSF under Grants DMR-9222812 and NSF-INT-9016257 and
M.P.N. by DMR-9214669 and CHE-9203498.  This research was conducted in
part using the resources of the Cornell Theory Center, which receives
major funding from the National Science Foundation (NSF) and New York
State, with additional support from the Advanced Research Projects
Agency (ARPA), the National Center for Research Resources at the
National Institutes of Health (NIH), IBM Corporation, and other members
of the center's Corporate Research Institute.

\newpage

\centerline{Table Captions}
\begin{enumerate}
\item
Critical exponents associated with the variables for which the
boundary conditions were twisted.
\item
Estimates of the central charge $c$ assuming the free energy per site
to be of the form $f_\infty+{\pi c \over 6 L^2}+{\alpha \over L^{2.8}}
+{\beta \over L^{3.8}}$.  Fits were made using data for
$L,L+2,\dots,30$.  For $L\ge 10$, $\beta$ was fixed at the value
obtained from the $L=4$ fit.  In all cases the normalized $\chi^2$ was
of order unity.
\end{enumerate}

\centerline{Figure Captions}

%================================================================
\begin{enumerate}

\item
Phase diagram of the $XY$-Ising model \protect{\cite{gkln91,lgk91}}.
Solid and dotted lines
indicate continuous and first-order transitions, respectively.
Filled circles indicate the locations where the present calculations
were performed.

\item Left: graphical representation of the conditional partition
function of a semi-infinite strip with helical boundary conditions,
i.e., the left eigenvector of the transfer matrix, which is shown on
the right.  In the lattice on the left, open circles represent sites
with variables $t_i$ and $\bf\rm n_i$ ($i=1,\dots,L$) that specify the
surface configuration upon which the conditional partition function
depends.  The full circles represent sites with variables that have
been summed over.  Right: graphical representation of the transfer
matrix. The variables associated with the circles make up the left
index of the matrix; the dots go with the right index.  Coincidence of
a circle and a dot produces a product of two $\delta$-functions.

\item Scaling plot of the interfacial free energy with Ising-twisted
boundary conditions; $A\approx 1$ is varied at constant $C=0.2885$.
\item Scaling plot of the interfacial free energy with $XY$-twisted
boundary conditions; $A\approx 1$ is varied at constant $C=0.2885$.

\item Scaling plot of the interfacial free energy with Ising-twisted
boundary conditions; $C\approx 1.32$ is varied at constant $A=2$.

\item Scaling plot of the interfacial free energy with $XY$-twisted
boundary conditions; $C\approx 1.32$ is varied at constant $A=2$.

\item Effective critical couplings $A_{\rm c}$ vs. $1/L$ and the
results of extrapolation to $L=\infty$ at $C=0.2885$ for both boundary
conditions.

\item Effective critical coupling $C_{\rm c}$ vs. $1/L$ and the results
of extrapolation to $L=\infty$ at $A=2$ for both boundary conditions.

\item Effective $y_T$ vs. $1/L$ for critical point at $A\approx 1$.

\item Effective $y_T$ vs. $1/L$ for critical point at $A=2$.

\item Effective $y^{(\rm d)}$ vs. $1/L$ for critical point at $A\approx
1$.

\item Effective $y^{(\rm d)}$ vs. $1/L$ for critical point at $A=2$.

\item Effective conformal anomaly vs inverse system size $1/L$
for various values of $A$.

\end{enumerate}
%====================================================================

\begin{table}[h] \caption{Critical exponents associated with the
variables for which the boundary conditions were twisted.}

\begin{tabular}{|l|l|l|}
		&Ising		&$XY$          \\
\tableline
$A\approx 1$	&$y_T=1.27$	&$y_T=0.97$	\\
		&$y_{\rm I}^{(\rm d)}=1.798$	&$y_{XY,{1 \over 2}}^{(\rm d)}=1.715$	\\
\tableline
$A=2$		&$y_T=1.45$	&$y_T=1.12$	\\
		&$y_{\rm I}^{(\rm d)}=1.801$	&$y_{XY,{1 \over 2}}^{(\rm d)}=1.616$	\\
\end{tabular}
\label{tab.exponents}
\end{table}

\begin{table}

\caption{Estimates of the central charge $c$ assuming the free energy
per site to be of the form $f_\infty+{\pi c \over 6 L^2}+{\alpha \over
L^{2.8}} +{\beta \over L^{3.8}}$.  Fits were made using data for
$L,L+2,\dots,30$.  For $L\ge 10$, $\beta$ was fixed at the value
obtained from the $L=4$ fit.  In all cases the normalized $\chi^2$ was
of order unity.}

\begin{tabular}{|c|c|c|}
                   & $L$                        & $c $      \\
\tableline
                   & 4                          & $1.466(6)$\\
                   & 6                          & $1.47(1) $\\
$A\approx 1$       & 8                          & $1.44(4) $\\
                   &10                          & $1.46(1) $\\
                   &12                          & $1.46(2) $\\
\tableline
                   & 4                          & $1.62(2) $\\
                   & 6                          & $1.57(4) $\\
$A = 2$            & 8                          & $1.48(9) $\\
                   &10                          & $1.57(3) $\\
                   &12                          & $1.56(6) $\\
\end{tabular}
\label{tab.central}
\end{table}

\begin{figure}
\begin{center}

%# 1 "transfer_matrix.tex"
%\documentstyle{article}
%\begin{document}
%\begin{figure}
\setlength{\unitlength}{2pt}
\begin{picture}(20,95)(-35,-20)
\put( 0,30){\circle{4}}
\put(-20,30){$t_L,{\rm\bf n}_L$}
\put( 0,40){\circle{4}}
\put( 0,40){\circle*{2}}
\put(-49,40){$s_{L},{\rm\bf m}_{L}\,|\,t_{L-1},{\rm\bf n}_{L-1}$}
\put(-1,48){\vdots}
\put( 0,60){\circle{4}}
\put( 0,60){\circle*{2}}
\put(10, 0){\circle{4}}
\put(10, 0){\circle*{2}}
\put(10,-10){\circle{4}}
\put(10,-10){\circle*{2}}
\put( 9, 8){\vdots}
\put(10,20){\circle{4}}
\put(10,20){\circle*{2}}
\put(15,20){$s_2,{\rm\bf m}_2\,|\,t_1,{\rm\bf n}_1$}
\put(10,30){\circle*{2}}
\put(15,30){$s_1,{\rm\bf m}_1$}
\put( 0,32){\line(0,7)7}
\put( 2,30){\line(7,0)7}
%lattice
\put(-80,50){\circle*{2}}
\put(-80,60){\circle*{2}}
\put(-81,38){\vdots}
\multiput(-80,30)(0,-10){3}{\circle*{2}}
\put(-81,-2){\vdots}
\put(-80,-10){\circle*{2}}

\put(-70,60){\circle{2}}
\put(-70,50){\circle{2}}
\put(-71,38){\vdots}
\put(-70,30){\circle{2}}
\multiput(-70,20)(0,-10){2}{\circle*{2}}
\put(-71,-2){\vdots}
\put(-70,-10){\circle*{2}}

\multiput(-60,20)(0,-10){2}{\circle{2}}
\put(-61,-2){\vdots}
\put(-60,-10){\circle{2}}

\put(-70,11){\line(0,8)8}
\put(-70,21){\line(0,8)8}
\put(-69,10){\line(8,0)8}

\put(-90,60){\dots}
\put(-90,50){\dots}
\multiput(-90,30)(0,-10){3}{\dots}
\put(-90,-10){\dots}

\put(-65,30){$t_L,{\rm\bf n}_L$}
\put(-55,20){$t_1,{\rm\bf n}_1$}
\put(-55,10){$t_2,{\rm\bf n}_2$}
\end{picture}

%\caption{Left: graphical representation of the conditional partition
%function of a semi-infinite strip with helical boundary conditions,
%i.e., the left eigenvector of the
%transfer matrix, which is shown on the right.  In the lattice on the
%left, open circles represent sites with variables $t_i$ and $\bf\rm
%n_i$ ($i=1,\dots,L$) that specify the surface configuration upon which
%the conditional partition function depends.  The full circles represent
%sites with variables that have been summed over.  Right: graphical
%representation of the transfer matrix. The variables associated with
%the circles make up the left index of the matrix; the dots go with the
%right index.  Coincidence of a circle and a dot produces a product of
%two $\delta$-functions.}
%
%\label{fig.transfer_matrix}
%\end{figure}
%\end{document}
%# 913 "paper_050595.acpp"

\caption{Left: graphical representation of the conditional partition
function of a semi-infinite strip with helical boundary conditions,
i.e., the left eigenvector of the transfer matrix, which is shown on
the right.  In the lattice on the left, open circles represent sites
with variables $t_i$ and $\bf\rm n_i$ ($i=1,\dots,L$) that specify the
surface configuration upon which the conditional partition function
depends.  The full circles represent sites with variables that have
been summed over.  Right: graphical representation of the transfer
matrix. The variables associated with the circles make up the left
index of the matrix; the dots go with the right index.  Coincidence of
a circle and a dot produces a product of two $\delta$-functions.}

\label{fig.transfer_matrix}
\end{center}
\end{figure}

\begin{figure}
\begin{center}

%# 1 "ScalingPlotxia1.istwist.tex"
% GNUPLOT: LaTeX picture
\setlength{\unitlength}{0.240900pt}
\ifx\plotpoint\undefined\newsavebox{\plotpoint}\fi
\sbox{\plotpoint}{\rule[-0.200pt]{0.400pt}{0.400pt}}%

%# 937 "paper_050595.acpp"

\caption{Scaling plot of the interfacial free energy with Ising-twisted
boundary conditions; $A\approx 1$ is varied at constant $C=0.2885$.}

\label{ScalingPlotxia1.istwist}
\end{center}
\end{figure}

\begin{figure}
\begin{center}

%# 1 "ScalingPlotxia1.xytwist.tex"
% GNUPLOT: LaTeX picture
\setlength{\unitlength}{0.240900pt}
\ifx\plotpoint\undefined\newsavebox{\plotpoint}\fi
\sbox{\plotpoint}{\rule[-0.200pt]{0.400pt}{0.400pt}}%
%
%# 952 "paper_050595.acpp"

\caption{Scaling plot of the interfacial free energy with $XY$-twisted
boundary conditions; $A\approx 1$ is varied at constant $C=0.2885$.}

\label{ScalingPlotxia1.xytwist}
\end{center}
\end{figure}

\begin{figure}
\begin{center}

%# 1 "ScalingPlotxia2.istwist.tex"
% GNUPLOT: LaTeX picture
\setlength{\unitlength}{0.240900pt}
\ifx\plotpoint\undefined\newsavebox{\plotpoint}\fi
\sbox{\plotpoint}{\rule[-0.200pt]{0.400pt}{0.400pt}}%
%
%# 967 "paper_050595.acpp"

\caption{Scaling plot of the interfacial free energy with Ising-twisted
boundary conditions; $C\approx 1.32$ is varied at constant $A=2$.}

\label{ScalingPlotxia2.istwist}
\end{center}
\end{figure}

\begin{figure}
\begin{center}

%# 1 "ScalingPlotxia2.xytwist.tex"
% GNUPLOT: LaTeX picture
\setlength{\unitlength}{0.240900pt}
\ifx\plotpoint\undefined\newsavebox{\plotpoint}\fi
\sbox{\plotpoint}{\rule[-0.200pt]{0.400pt}{0.400pt}}%
%
%# 982 "paper_050595.acpp"

\caption{Scaling plot of the interfacial free energy with $XY$-twisted
boundary conditions; $C\approx 1.32$ is varied at constant $A=2$.}

\label{ScalingPlotxia2.xytwist}
\end{center}
\end{figure}

\begin{figure}
\begin{center}

%# 1 "xia1.Kc_L.tex"
% GNUPLOT: LaTeX picture
\setlength{\unitlength}{0.240900pt}
\ifx\plotpoint\undefined\newsavebox{\plotpoint}\fi
\sbox{\plotpoint}{\rule[-0.200pt]{0.400pt}{0.400pt}}%
%
%# 997 "paper_050595.acpp"

\caption{Effective critical couplings $A_{\rm c}$ vs. $1/L$ and the
results of extrapolation to $L=\infty$ at $C=0.2885$ for both boundary
conditions.}

\label{xia1.Kc_L}
\end{center}
\end{figure}

\begin{figure}
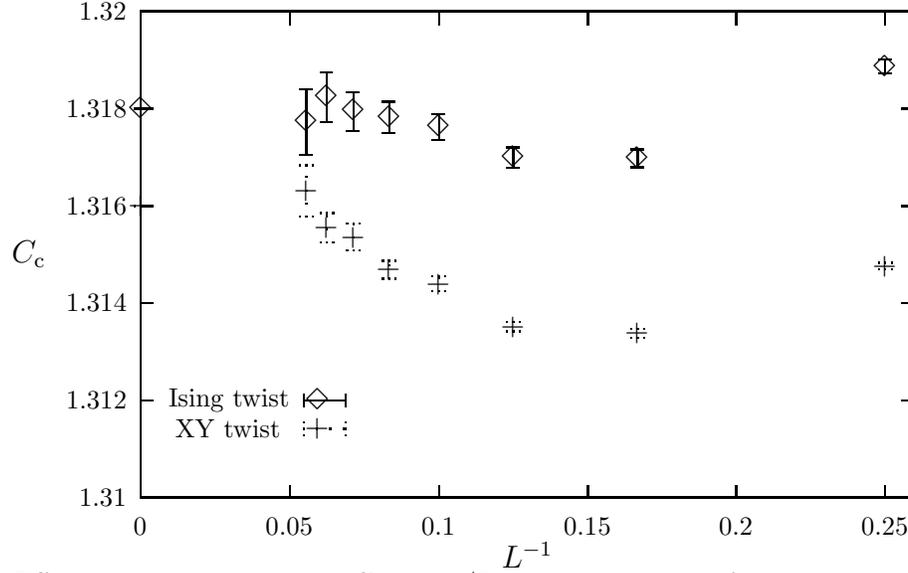

\begin{center}

%# 1 "xia2.Kc_L.tex"
% GNUPLOT: LaTeX picture
\setlength{\unitlength}{0.240900pt}
\ifx\plotpoint\undefined\newsavebox{\plotpoint}\fi
\sbox{\plotpoint}{\rule[-0.200pt]{0.400pt}{0.400pt}}%
%
%# 1013 "paper_050595.acpp"

\caption{Effective critical coupling $C_{\rm c}$ vs. $1/L$ and
the results of extrapolation to $L=\infty$ at $A=2$
for both boundary conditions.}
\label{xia2.Kc_L}
\end{center}
\end{figure}

\begin{figure}
\begin{center}

%# 1 "xia1.yt_L.tex"
% GNUPLOT: LaTeX picture
\setlength{\unitlength}{0.240900pt}
\ifx\plotpoint\undefined\newsavebox{\plotpoint}\fi
\sbox{\plotpoint}{\rule[-0.200pt]{0.400pt}{0.400pt}}%
%
%# 1027 "paper_050595.acpp"

\caption{Effective $y_T$ vs. $1/L$ for critical point at $A\approx 1$.}
\label{xia1.yt_L}
\end{center}
\end{figure}

\begin{figure}
\begin{center}

%# 1 "xia2.yt_L.tex"
% GNUPLOT: LaTeX picture
\setlength{\unitlength}{0.240900pt}
\ifx\plotpoint\undefined\newsavebox{\plotpoint}\fi
\sbox{\plotpoint}{\rule[-0.200pt]{0.400pt}{0.400pt}}%
%
%# 1039 "paper_050595.acpp"

\caption{Effective $y_T$ vs. $1/L$ for critical point at $A=2$.}
\label{xia2.yt_L}
\end{center}
\end{figure}

\begin{figure}
\begin{center}

%# 1 "xia1.yh_L.tex"
% GNUPLOT: LaTeX picture
\setlength{\unitlength}{0.240900pt}
\ifx\plotpoint\undefined\newsavebox{\plotpoint}\fi
\sbox{\plotpoint}{\rule[-0.200pt]{0.400pt}{0.400pt}}%
%
%# 1051 "paper_050595.acpp"

\caption{Effective $y^{(\rm d)}$ vs. $1/L$ for critical point at $A\approx 1$.}
\label{xia1.yh_L}
\end{center}
\end{figure}

\begin{figure}
\begin{center}

%# 1 "xia2.yh_L.tex"
% GNUPLOT: LaTeX picture
\setlength{\unitlength}{0.240900pt}
\ifx\plotpoint\undefined\newsavebox{\plotpoint}\fi
\sbox{\plotpoint}{\rule[-0.200pt]{0.400pt}{0.400pt}}%
%
%# 1063 "paper_050595.acpp"

\caption{Effective $y^{(\rm d)}$ vs. $1/L$ for critical point at $A=2$.}
\label{xia2.yh_L}
\end{center}
\end{figure}

\begin{figure}
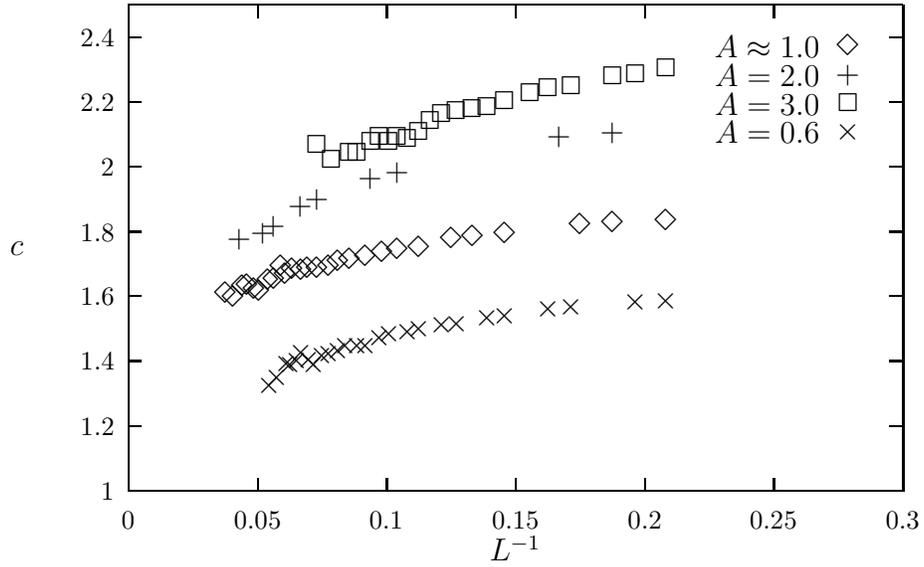

\begin{center}

%# 1 "plot.tex"
% GNUPLOT: LaTeX picture
\setlength{\unitlength}{0.240900pt}
\ifx\plotpoint\undefined\newsavebox{\plotpoint}\fi
\sbox{\plotpoint}{\rule[-0.200pt]{0.400pt}{0.400pt}}%
%
%# 1075 "paper_050595.acpp"

\caption{Effective conformal anomaly vs inverse system size $1/L$ for
various values of $A$}
\label{c.plot}
\end{center}
\end{figure}

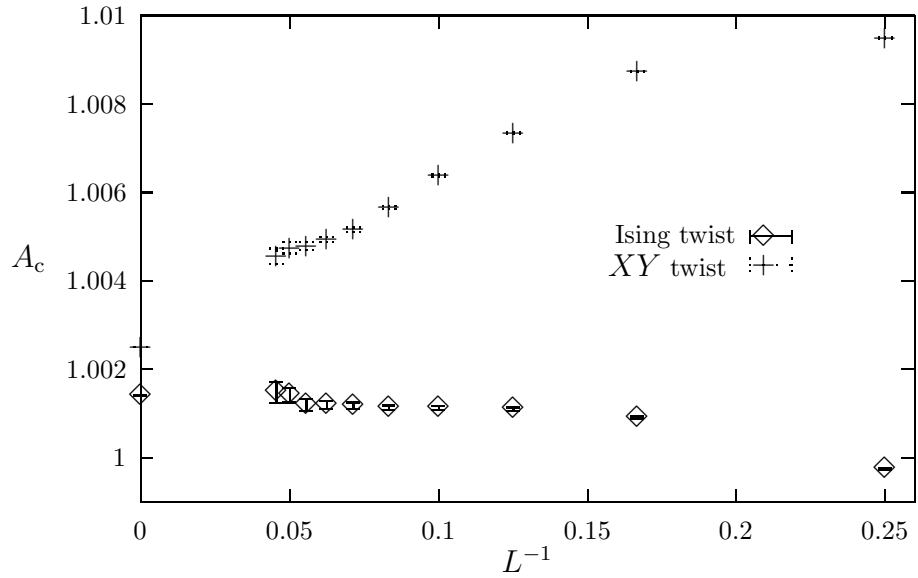
\end{document}